
%
%
\normalbaselineskip=12pt
\baselineskip=12pt
\magnification=1200
\hsize 15.0truecm 
\vsize 22.0truecm 
\nopagenumbers
\headline={\ifnum \pageno=1 \hfil \else\hss\tenrm\folio\hss\fi}
\pageno=1

\def\lsim{\mathrel{\rlap{\lower4pt\hbox{\hskip1pt$\sim$}}
    \raise1pt\hbox{$<$}}}         
\def\gsim{\mathrel{\rlap{\lower4pt\hbox{\hskip1pt$\sim$}}
    \raise1pt\hbox{$>$}}}         


\centerline{\bf An interesting charmonium state formation and decay:
$p\bar p \to$ $^1D_2 \to$ $^1P_1 \gamma$}
\vskip 36pt
\centerline{M. Anselmino$^a$, F. Caruso$^{b,c}$, F. Murgia$^d$
and M.R. Negr\~ao$^b$}
\vskip 12pt
\centerline{\it {}$^a$Dipartimento di Fisica Teorica, Universit\`a di
 Torino}
\centerline{\it and Istituto Nazionale di Fisica Nucleare,
Sezione di Torino}
\centerline{\it Via Pietro Giuria 1, I--10125 Torino, Italy}
\vskip 6pt
\centerline{\it {}$^b$Centro Brasileiro de Pesquisas F\'\i sicas/CNPq}
\centerline{\it Rua Dr. Xavier Sigaud 150, 22290-180,
Rio de Janeiro, RJ, Brazil}
\vskip 6pt
\centerline{\it {}$^c$Instituto de F\'\i sica, Universidade do Estado do
Rio de Janeiro, Brazil}
\centerline{\it Rua Sao Francisco Xavier 524, 20550--011,
Rio de Janeiro, RJ, Brazil}
\vskip 6pt
\centerline{\it {}$^d$Istituto Nazionale di Fisica Nucleare,
Sezione di Cagliari}
\centerline{\it Via A. Negri 18, I--09127 Cagliari, Italy}
\vskip 1.0in
\centerline{\bf ABSTRACT}
\vskip 12pt
Massless perturbative QCD forbids, at leading order, the exclusive
annihilation of proton-antiproton into some charmonium states, which,
however, have been observed in the $p\bar p$ channel, indicating the
significance of higher order and non perturbative effects in the few
GeV energy region. The most well known cases are those of the
$^1S_0$ ($\eta_c$) and the $^1P_1$. The case of the $^1D_2$ is
considered here and a way of detecting such a state through its typical
angular distribution in the radiative decay $^1D_2 \to$ $^1P_1 \gamma$
is suggested. Estimates of the branching ratio $BR(^1D_2 \to p\bar p)$,
as given by a quark-diquark model of the nucleon, mass corrections and
an instanton induced process are presented.
\vskip 36pt
\rightline {\hfill CBPF-NF-014/94}
\rightline {\hfill INFNCA-TH-94-5}
\rightline {\hfill January 1994}

\vfill
\eject
\baselineskip 18pt plus 2pt minus 2pt

         Several hadronic two-body decays of charmonium states are
forbidden within leading order perturbative QCD [1,2], but are
nevertheless observed to occur with decay rates comparable to or even
bigger than those of allowed decays [3]; the most well known examples
are the $J/\psi \to VP$ [4], and the $\eta_c \to B\bar B, \, VV$ [5]
channels, where $P$ is a psdeudoscalar meson, $V$ a vector meson and
$B$ a baryon. Indeed the observed decay rates for $J/\psi \to \rho \pi,
\, K^* \bar K$  and $\eta_c \to p\bar p, \, \rho\rho, \, \phi\phi, \,
K^* \bar K^*$ are difficult to explain within conventional perturbative
QCD. Recently, also the $^1P_1$ coupling to $p\bar p$ has been
established [6], despite being equally forbidden by the helicity
conservation rule of massless pQCD [7].

         Among the attempts to solve these problems, non leading
contributions [8], two quark correlations inside baryons [9], quark
mass effects [10] and gluonic contents of the $J/\psi$ [4] and the
$\eta_c$ [11] have been considered. Higher order Fock states might
help with the $J/\psi \to \rho \pi$ decay [8], but their contributions
to other processes are not clear;
diquarks and mass corrections do not help much with the $\eta_c$
forbidden decays, whereas gluonic contributions seem to be more
promising [12]. Recently a dynamical model for such contributions,
with instanton-induced non perturbative chiral symmetry breaking, has
been used to obtain a good agreement with the data on $\Gamma(\eta_c
\to p\bar p)$ [13].

         We consider here yet another case of a charmonium decay which
should be forbidden according to perturbative QCD, namely $^1D_2 \to
p\bar p$. Its observation would be very interesting because, among the
above non perturbative mechanisms invoked to explain the other forbidden
decays, no one seems to be able to account for a sizeable decay rate:
as we shall see, both mass corrections and diquarks give very tiny decay
rate values and instanton induced processes are strongly suppressed with
increasing $Q^2$ values [13].

         Let us briefly recall why the $^1D_2 \to p\bar p$ decay is
forbidden in massless pQCD. This charmonium state has quantum numbers
$J^{PC} = 2^{-+}$: parity, angular momentum and charge conjugation
conservation only allow a final $p\bar p$ state with orbital angular
momentum $L=2$ and spin $S=0$. $S=0$ implies that the $p$ and
the $\bar p$ must have, in the charmonium rest frame,
the same helicity, which is forbidden by the QCD vector coupling
of hard gluons to massless quarks and antiquarks. Such helicity selection
rule can only be broken by terms proportional to $m_q/m_c$ or $k_T/m_c$,
where $m_q$ and $m_c$ are respectively the light quark and charmed
quark masses and $k_T$ is the quark intrinsic transverse momentum. The
current masses of $u$ and $d$ quarks are very small compared to the
charmed quark mass and terms proportional to $m_q/m_c$ are indeed
negligible; terms proportional to $k_T/m_c$ might be more relevant,
but no comprehensive treatment of these contributions, together with
other higher twist effects, has yet been performed.

         Let us now consider the $^1D_2$ state created in $p\bar p$
annihilations, choosing the $z$-axis as the proton direction in the
$p\bar p$ centre of mass frame. It is then clear, from what we said
above, that the $^1D_2$ state can only be created with the spin third
component $J_z=0$; such charmonium state is then purely polarized and
its spin density matrix has only one non zero component,
$$\rho_{00}(^1D_2) = 1 \,. \eqno(1)$$

         This peculiar property reflects into the decay angular
distributions of the $^1D_2$. One radiative decay which is expected
to be observed with a large branching ratio is
$$^1D_2 \to {}^1P_1 \gamma \,, \eqno(2)$$
which is dominated by an electric dipole transition. The angular
distribution of the photon, as it emerges in the rest frame of the
$^1D_2$, is then simply given by [14]
$$W_{\gamma}(\theta) = {1\over 8}(5-3\cos^2\theta) \,, \eqno(3)$$
where $\theta$ is the photon polar angle and an integration has been
performed over the azimuthal angle.

         The observation of such an angular distribution in $p\bar p$
exclusive annihilations should then be a clear signal of the formation
and decay of the $^1D_2$ state; the full chain of processes to be
looked for, according to the observed or expected decays of the
$^1P_1$ state [6], is:
$$p\bar p \to {}^1D_2 \to {}^1P_1 \gamma \to (\eta_c \gamma) \, \gamma
\to (\gamma\gamma\gamma) \, \gamma \,, \eqno(4)$$
or
$$p\bar p \to {}^1D_2 \to {}^1P_1 \gamma \to (J/\psi \pi^0) \, \gamma
\to (e^+e^- \pi^0) \, \gamma \,. \eqno(5)$$
Notice that the expected mass of the $^1D_2$ state is
$M_{_D} = (3788 \pm 7)$ MeV [15].

         So far the $^1S_0 \, (\eta_c)$, $^3S_1 \, (J/\psi$ and
$J^\prime)$, $^3P_1 \, (\chi_{c1})$ and $^3P_2 \, (\chi_{c2})$ charmonium
states have been observed to couple to $p\bar p$; the corresponding
branching ratios, $BR(^{2S+1}L_J \to p\bar p)$, are typically
of the order of $10^{-4}$ to $10^{-3}$ [3]. Curiously, the $\eta_c \to
p\bar p$ branching ratio, which should be zero according to lowest order
perturbative QCD, is among the largest ones. Recently, also the $^1P_1$
has been observed in the $p\bar p \to {}^1P_1 \to J/\psi \pi^0$ channel
[6], with an estimate for the product of the two branching ratios
$BR(^1P_1 \to p\bar p)\,BR(^1P_1 \to J/\psi \pi^0)$ $\simeq 10^{-7}$.
Notice that, similarly to what explained for the $^1D_2$, also the
$^1P_1$ decay into $p\bar p$ is forbidden by leading order pQCD [7].
The $^3P_0$ state has not yet been observed, but this is
presumably due to its small ($\lsim 10^{-2}$) branching ratio into
$J/\psi \,\gamma$; this makes the full process through which one looks
for such a state, $p\bar p \to \chi_{c0} \to J/\psi \gamma \to
e^+e^-\gamma$, a difficult one to detect. The analogous situation for
the $^3P_1$ and $^3P_2$ states is much more favourable in that their
branching ratios into $J/\psi \gamma$ are respectively $\simeq 0.27$
and 0.13 [3].

         Thus, it is natural to expect a $^1D_2 \to p\bar p$ branching
ratio similar to that observed for the other charmonium states. However,
this would be very difficult to explain; to see why, we now briefly
consider several possible non perturbative contributions to such
process.

Mass corrections to `forbidden' charmonium decays have been considered
in Ref. [10] for $\eta_c, \, \chi_{c0} \to p\bar p$; they
yield sizeable values of $\Gamma(\chi_{c0} \to p\bar p)$, but very small
ones for $\Gamma(\eta_c \to p\bar p)$, actually a factor $\sim 10^{-4}$
smaller than data. Following the same procedure and notations as in
Ref. [10] we have computed the helicity amplitudes for the decay
$^1D_2 \to p\bar p$; the only non zero ones are:
$$\eqalign{A_{++;M}(\theta) &= -A_{--;M}(\theta) = {32\over 27}
\sqrt{{5\over 3}} \pi^4 \alpha_s^3 R^{\prime\prime}(0)
{F_N^2\over M_{_D}^7} \, \epsilon \, (1-4\epsilon^2)^2 \cr
&d^2_{M,0}(\theta)
\int_0^1 \, dx_2 \int_0^{1-x_2} \, dx_3
\int_0^1 \, dy_2 \int_0^{1-y_2} \, dy_3 \cr
&{1\over [x_2y_2 + (x_2-y_2)^2\epsilon^2]}
{1\over [1 + x_2y_2 - x_2 - y_2 + (x_2-y_2)^2\epsilon^2]} \cr
& {1\over [x_3y_3 + (x_3-y_3)^2\epsilon^2]}
  {1\over [ (1-x_2)y_3 + (1-x_2-y_3)^2 \epsilon^2]} \cr
& {1\over \bigl[ x_2y_2 - {1\over 2}(x_2+y_2) + (x_2-y_2)^2
\epsilon^2 \bigr]^3} \, (x_2 - y_2)^3 \times \cr
&\Biggl\{ - \Bigl[ \varphi_x(231)\varphi_y(321) - \varphi_x(132) \bigl[
\varphi_y(321)+\varphi_y(123) \bigr] \cr
& \quad\quad\quad\quad\quad\quad
-  \bigl[ \varphi_x(132)+\varphi_x(231)
\bigr]\varphi_y(123) \Bigr] (1 - x_2 - y_3) \cr
&-\Bigl[ \varphi_x(123)\varphi_y(213) - \varphi_x(321) \bigl[
\varphi_y(213)+\varphi_y(312) \bigr] \cr
& \quad\quad\quad\quad\quad\quad
- \bigl[ \varphi_x(321)+\varphi_x(123)
\bigr]\varphi_y(312)  \Bigr] (1 - x_2) \cr
&+\Bigl[ \varphi_x(213)\varphi_y(123) - \varphi_x(312) \bigl[
\varphi_y(321)+\varphi_y(123) \bigr] \cr
& \quad\quad\quad\quad\quad\quad
- \bigl[ \varphi_x(213)+\varphi_x(312)
\bigr]\varphi_y(321)  \Bigr] (1 - x_2) \Biggr\} \cr} \eqno(6)$$
where $M_{_D}$ is the $^1D_2$ mass and $R^{\prime\prime}(0)$ is the value at
the origin of its wave function second derivative. $\varphi_z(i,j,k)
\equiv \varphi(z_i,z_j,z_k)$ denotes
the proton distribution amplitude and we refer to Ref. [10] for further
details. Here we only notice that $\epsilon$ is the ratio of the proton
to the charmonium mass, $\epsilon=m_p/M_{_D}$, so that in the massless
limit, $\epsilon \to 0$, indeed $A_{\pm\pm;M} = 0$, as required by
perturbative QCD.

         From Eq. (6) one obtains the decay rate
$$\eqalign{\Gamma(^1D_2 \to p\bar p) &=
{(1-4\epsilon^2)^{1/2}\over 40\,(2\pi)^4}
\sum_{\lambda_p,\lambda_{\bar p},M}\,
\int_{-1}^1 d(\cos\theta)
\, \vert A_{\lambda_p,\lambda_{\bar p},M}(\theta) \vert^2 \cr
&={2^5\over 3^7} \pi^4 \alpha_s^6 \vert
R^{\prime\prime}(0) \vert^2 \vert F_N \vert^4 \epsilon^2
(1-4\epsilon^2)^{9/2} {I^2(\epsilon) \over M_{_D}^{14}} \cr} \eqno(7)$$
where $I$ is the multiple integral appearing in Eq. (6).

         From Ref. [9]\footnote*
{Notice that in Ref. [9] the $^1D_2$ state is named $f_2$}
also the decay rate of the $^1D_2$ into two
gluons can be obtained:
$$\Gamma(^1D_2 \to gg) = {32\over 3} {\alpha_s^2 \over M_{_D}^6} \vert
R^{\prime\prime}(0) \vert^2 \,. \eqno(8)$$
By assuming the total hadronic decay rate of the $^1D_2$ to be
approximately given by Eq. (8), one can get an estimate of the branching
ratio $BR(^1D_2 \to p\bar p)$ by taking the ratio of Eqs. (7) and (8),
so that the unknown quantity $R^{\prime\prime}(0)$ cancels out. The
result strongly depends on the choice of the distribution amplitudes
$\varphi(x_1,x_2,x_3)$; according to the
different choices adopted in Ref. [10] one obtains
$$BR(^1D_2 \to p\bar p) \sim 10^{-8} \div 10^{-12} \,. \eqno(9)$$

         Eq. (9) clearly shows how mass corrections could not account
for the eventual observation of the $^1D_2 \to p\bar p$ decay; the
small values obtained for the branching ratio are mainly due to the
factor $(x_2-y_2)^3$ contained in the decay amplitude, Eq. (6). This is
similar to what happens for the $\eta_c \to p\bar p$ process, where mass
corrections are also very small, due to a factor $(x_2-y_2)$ in the
amplitude [10]; in the present case, actually, the situation is even
worse, because of the third power of $(x_2-y_2)$. In fact, in the
$\eta_c$ case, mass corrections lead to $BR(\eta_c \to p\bar p)
\sim 10^{-6} \div 10^{-10}$ [10], a result far away from the observed
value $BR(\eta_c \to p\bar p) \simeq 10^{-3}$, but bigger than
the values given in Eq. (9).

         One can similarly show that also two quark correlations
could not explain a branching ratio for the $^1D_2 \to p\bar p$ decay
of the order of $10^{-4}$; a vector diquark component of the proton
allows the decay, by allowing helicity flips at the gluon-vector
diquark coupling [9], but, once more, the numerical values turn out to
be too small. This can be explicitly checked by repeating the same
procedure followed above for mass corrections; the expression of the
decay helicity amplitudes, in the quark-diquark model of the proton,
can be found in Ref. [9]
and, again, it contains a small factor $(x-y)^3$. One finds, with
little dependence on the choice of the distribution amplitudes,
$$BR(^1D_2 \to p\bar p) \sim 10^{-8} \,. \eqno(10)$$

         Among other non perturbative effects proposed to explain
unexpectedly large branching ratios, the presence of the fundamental
($L=0$) trigluonium states, with quantum numbers $J^{PC} = 0^{-+}, \,
1^{--}, \, 3^{--}$, in the 3 GeV mass region, has been proposed
[4,11]. The first two states, mixing respectively with the $\eta_c$
and the $J/\psi$, might explain some of their `mysterious' decays.
However, a similar explanation for the $^1D_2$, the presence of a
$2^{-+}$ glueball with a mass close to 3.8 GeV, looks much less
natural and realistic.

         Let us consider finally the instanton induced mechanism
proposed in Ref. [13] for the $\eta_c \to p\bar p$ decay: we know
that its contribution decreases very rapidly with increasing $Q^2$
and, indeed, already for the decay of $\eta_c^\prime$, with a mass
$\simeq 3.6$ GeV, is much smaller than for the $\eta_c$ [13].
Considering the still higher mass of the $^1D_2$, $M_{_D} \simeq 3.8$ GeV,
we cannot expect this non perturbative contribution to be large
enough to produce a branching ratio for the process $^1D_2 \to
p\bar p$ similar to those observed for the other charmonium states.

         We have thus seen how several possible non perturbative
effects cannot contribute significantly to the $^1D_2$
coupling to $p\bar p$; on the other hand we know that leading order
pQCD predicts no coupling at all, whereas higher order corrections
are difficult to evaluate and have never been computed. A similar
situation occurs with the $\eta_c$, with the difference that for such
particle one might expect a significant gluonic contribution [12,13].
Therefore, the eventual observation of a $BR(^1D_2 \to p\bar p)
\sim 10^{-4}$, analogous to the values observed for all other
charmonium states which can couple to $p\bar p$, would pose an
intriguing challenge to the theory.

          The $^1D_2$ state could be looked for in the mass region
$M_{_D} \simeq 3788$ MeV [15] and in the reactions suggested by Eqs. (4)
and (5), which should exhibit a typical decay angular distribution
of the $\gamma$ in the first step of the process. In fact the $^1D_2$
created in $p\bar p$ annihilation is in a pure spin state with $J_z=0$
and its decay into $^1P_1 \gamma$, dominated by an $E1$ transition,
has the simple angular distribution given in Eq. (3). Hopefully, such
radiative decay has a large branching ratio, so that the processes of
Eqs. (4) and/or (5) can be detected. This is not unrealistic if one
notices that the $^1D_2$ state, due to its expected mass and quantum
numbers, cannot decay into pairs of $D$ and/or $D^*$ mesons.

\vskip 24pt
\noindent
{\bf Acknowledgements}

M.A. would like to thank the members of the LAFEX group at the CBPF
for their warm hospitality; F.C. is grateful to the CNPq of Brasil for
a research grant and M.R.N. is grateful to the CAPES of Brasil for a
fellowship. We thank R. Mussa for useful discussions.

\vfill\eject
\noindent {\bf References}
\item{[ 1]} S.J.~Brodsky and G.P.~Lepage; {\it Phys. Rev.} {\bf D24}
(1981) 2848
\item{[ 2]} V.L.~Chernyak and A.R.~Zhitnitsky, {\it Nucl. Phys.}
{\bf B201} (1982) 492
\item{[ 3]} K.~Hikasa {\it et al.}, {\it Phys. Rev.} {\bf D45 II} (1992)
\item{[ 4]} S.J.~Brodsky, G.P.~Lepage and S.F.~Tuan, {\it Phys. Rev.
Lett.} {\bf 59} (1987) 621
\item{[ 5]} M.~Anselmino, F.~Caruso and F.~Murgia, {\it Phys. Rev.}
{\bf D42} (1990) 3218
\item{[ 6]} T.A.~Armstrong {\it et al.}, {\it Phys. Rev. Lett.} {\bf
16} (1992) 2337
\item{[ 7]} A.~Andrikopoulou, {\it Z. Phys.} {\bf C22} (1984) 63
\item{[ 8]} V.L.~Chernyak and A.R.~Zhitnitsky, {\it Phys. Rep.}
{\bf 112} (1984) 173.
\item{[ 9]} M.~Anselmino, F.~Caruso and S.~Forte, {\it Phys. Rev.}
{\bf D44} (1991) 1438
\item{[10]} M.~Anselmino, R.~Cancelliere and F.~Murgia, {\it Phys.
Rev.} {\bf D46} (1992) 5049
\item{[11]} M.~Anselmino, M.~Genovese and E.~Predazzi, {\it Phys. Rev.}
{\bf D44} (1991) 1597
\item{[12]} M.~Anselmino, M.~Genovese and D.E.~Kharzeev, Torino
{\it preprint} DFTT 64/93 (1993)
\item{[13]} M.~Anselmino and S.~Forte, Torino {\it preprint} DFTT 73/93
(1993), to appear in {\it Phys. Lett.} {\bf B}
\item{[14]} See, {\it e. g.}, T.A.~Armstrong {\it et al.}, {\it Phys.
Rev.} {\bf D48} (1993) 3037; M.~Anselmino, F.~Caruso and R.~Mussa,
{\it Phys. Rev.} {\bf D45} (1992) 4340
\item{[15]} D.B.~Lichtenberg, R.~Roncaglia, J.G.~Wills, E.~Predazzi
and M.~Rosso, {\it Z. Phys.} {\bf C46} (1990) 75.

\bye